\documentclass{article}
\usepackage{authblk}
\usepackage{hyperref}
\usepackage{graphicx}

\providecommand{\keywords}[1]{\textbf{\textit{Keywords:}} #1}
\begin{document}
	
	
	\title{Discovering New Variable Stars at Key Stage 3}
	
	\author{Katy Chubb\footnote{
			katy.chubb.14@ucl.ac.uk}, Rosie Hood, Thomas Wilson, Jonathan Holdship,}
	
	\affil{Department of Physics and Astronomy, University College London,\\
		London, WC1E 6BT, UK\\
	}

	\author{and Sarah Hutton}

	\affil{The Henrietta Barnett School, Central Square, Hampstead 1-1 Garden Suburb, London NW11 7BN
	}

	\date{}

\maketitle

\begin{abstract}

Details of the London pilot of the `Discovery Project' are presented, where university-based astronomers were given the chance to pass on some real and applied knowledge of astronomy to a group of selected secondary school pupils. It was aimed at students in Key Stage 3 of their education, allowing them to be involved in real astronomical research at an early stage of their education, the chance to become the official discoverer of a new variable star, and to be listed in the International Variable Star Index database\cite{VSX_web}, all while learning and practising research-level skills. Future plans are discussed. \\

\end{abstract}


\section{Introduction}\label{sec:intro}

In this paper details of the London pilot of the Russian-based ``Discovery project'' are presented, where a group of students were given the chance to learn and implement all the skills required to discover a new variable star over a series of extra-curricular sessions. It offered numerous benefits to both astronomers and students, including the opportunity for the students to be listed as the official discover of a new variable star in the International Variable Star Index, VSX~\cite{VSX_web}. They were taught the basic skills and background knowledge before analysing recent data from telescopes primarily located in Australia and conducting university-level research. \\

The project was first developed and pioneered by the team at the Noosfera Education Support Foundation and professional astronomers at Sternberg Astronomical Institute, based in Moscow, Russia, where the course has been successfully running for almost two years~\cite{noosfera_web}. Since its launch there have been a number of discoveries of variable stars and asteroids made by participating students~\cite{discoveries_web}, which can be found on the VSX database, by searching by name, for example ``Kachalin 1''~\cite{kachalin_web}. In early 2016 the decision was made to expand the scheme and London was chosen as the first location, a collaboration between the Mayor\textsc{\char13}s Fund for London and Noosfera.\\

The first London ``Discovery'' Clubs were held over ten weekly sessions at Hammersmith and UCL Academies, with each consisting of around ten 11-12 year olds, in Key Stage 3 of their education. In recent years Ofsted has observed a lack of support for Key Stage 3 students, particularly those with pupil premium funding; more priority was found to be given to Key Stage 4 and 5 and there was found to be a lack of challenge for the most able students~\cite{ofsted_report_web}. Priority for the Discovery club was therefore given to pupil premium students; pupils from disadvantaged backgrounds as indicated by their eligibility for free school meals. Each club was run by a team of two PhD level astronomers from UCL, supported by one of the academy\textsc{\char13}s teachers and UCL\textsc{\char13}s Outreach Co-ordinator and Ogden Science Officer. As a result, a total of nine new variable stars have been discovered by eight Key Stage 3 students. Four of these have been submitted and accepted into the VSX database. \\

To date there have been several investigations into the benefits of bringing research into schools in order to educate and encourage students to get involved in science from a young age~\cite{00Hoxxxx,14FiHoRe,SEPnet_web}, along with additional benefits for the researchers; in particular improved communication skills~\cite{16ClRuEn,11HaHaCh}. A number of projects have been set up to that effect, such as the US-based International Astronomical Search Collaboration~\cite{iasc_web} and the UK-based Institute for Research in Schools (IRiS)~\cite{iris_web}, which facilitates school students and teachers to participate in university-based STEM research projects. Some other UK-based examples include the Brilliant Club~\cite{Brilliant_web}, Researchers in Schools\cite{ris_web} and EduTwinkle~\cite{edutwinkle_web}, the first education programme linked to a UK-based space mission. This includes ORBYTS (Original Research by Young Twinkle Students)~\cite{orbyts_web,16SoBaCh}, which partners young university-based scientific researchers with A-level students to perform original, peer-reviewed, university level research. The Cosmic ray detectives, who exhibited at the 2015 Royal Society Summer Science Exhibition~\cite{ss_web}, was a collaboration between researchers at the University of Birmingham and Bordersley Green Girls school and sixth form. It was formed as part of the HiSPARC international research network~\cite{05FrHeOo}, which collects data from cosmic ray detectors built by schools and installed on their roofs. The Discovery Project is a complementary addition to these studies, introducing students to scientific research at an earlier age.

\section{Methods and Software}

The method of discovery used was developed and pioneered by the ``Discovery'' team at the Moscow-based Noosfera Foundation~\cite{noosfera_web}, with the lessons plans altered slightly to fit the UK education system. \\

The sessions were planned with the intention of encouraging questions from and maximising interaction with the students, allowing them to be fully involved in the learning process. This involved a mixture of introducing basic astronomy concepts, holding general discussions and providing hands-on software tutorials, along with close support throughout. Access to a projector in order to demonstrate and present work aided in these efforts but was not necessarily a requirement. The project required assistance and collaboration with the school IT departments in order to ensure all software was correctly installed. Here a brief outline of the software required for the discovery process is given. \\

Where possible, use was made of free software. In particular, Stellarium~\cite{Stellar_web}, an open source planetarium software, proved very useful and intuitive for the students. It was employed to demonstrate basic astronomy concepts such as co-ordinate systems and magnitudes, which proved useful later on in the course when students were required to record the right ascension and declination of their target stars.\\ 

Astrometrica, also available for free download~\cite{astro_web}, is an interactive software tool used for asteroid detection. It can be linked to catalogues in order to determine if an asteroid has been previously discovered or not and has the capability to produce the relevant reports to submit an asteroid to the Minor Planet Centre. At least one session was dedicated to teaching the students about asteroid detection using this software. \\

\begin{figure}
	\centering
	\includegraphics[width=0.7\linewidth]{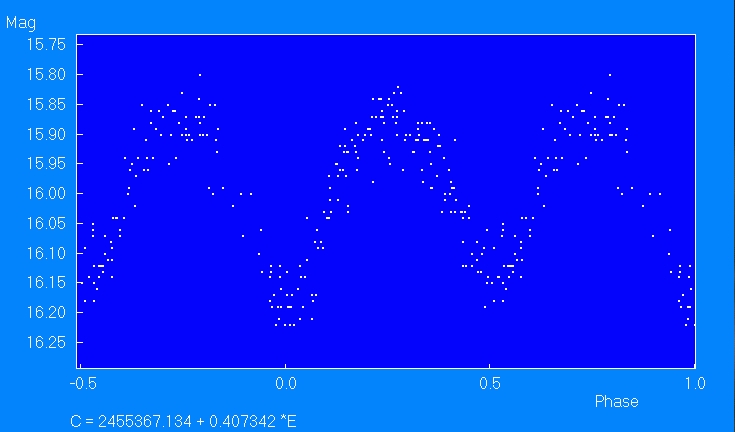}
	\caption{The lightcurve for Mousa v2, one of the variable stars discovered at Hammersmith Academy. 0.407342 is the period in days and 2455367.134 is the epoch of primary eclipse in Julian days (19 Jun 2010). The y-axis gives the magnitude, a measure of the stellar brightness.}
	\label{fig:mv2}
\end{figure}

A combination of Maxim DL~\cite{Maxim_web} and MuniWin~\cite{Muniwin_web}, examples of astronomical imaging software, were used to analyse images and generate light curves in order to detect variable stars. MuniWin, available free of charge, is ideal for analysing and outputting relevant light curve information (variation in a star\textsc{\char13}s brightness over time) while Maxim DL is required for finding each set of stellar coordinates. All photometric data reduction was completed by the astronomers prior to the astronomy sessions, which can be done using either package. Maxim DL is particularly user friendly, however it is only free for a trial period of 30 days. There is equivalent, though slightly less user friendly, software available such as SAOImage DS9~\cite{ds9_web} which can be downloaded free of charge or other more user friendly software packages which are suitable for photometry of this nature such as SalsaJ~\cite{salsaj_web} or Makali'i~\cite{makalii_web}. If the light curve generated using MuniWin appeared to show some variability then the students would search existing databases to check if the star had already been discovered. This was easily be carried out using the search function in the aforementioned online International Variable Star Index database~\cite{VSX_web}. If it had not been submitted to the database already then more detailed information on the light curve, from data collected over a larger timescale, could be found by searching for the coordinates of the star in the Catalina Sky Survey database~\cite{catalina_web}. This data could be downloaded and analysed in WinEffect~\cite{wineffect_web}, such as the example given in figure \ref{fig:mv2}. Similar software is available for free download, such as VARTOOLS~\cite{16HaBaxx}.\\

The team of astronomers would then work with the students to collect together all necessary data and create reports in the correct format for submission. These included coordinates, brightness and characterisation of the type of variable star found.\\

After being taught the necessary skills using images containing previously discovered stars or asteroids, the students were given new images in order to independently search for their own new discoveries. Such images can be requested from iTelescope.net~\cite{itel_web} for a subscription fee, or alternatively for free from UK-based projects such as the National Schools Observatory~\cite{NSO_web} and the Faulkes Telescope Project~\cite{faulkes_web}. Those used in the pilot were all taken from the T31 Deep Space telescope from the Sliding Spring Observatory in Coonabarabran, New South Wales, Australia.\\

A printed Young Astronomers Student Workbook was also provided for each student, translated and adapted from the equivalent Russian version. These allowed them to record their progress and contained valuable guidance and information, including short exercises and practical computational instructions. If the software failed, which happened on a few occasions, then the exercises in this booklet were a good alternative. \\

An official closing ceremony marked the club finales, set up and attended by members of the Noosfera team from Moscow. All participants were awarded certificates, trophies and goody bags.

\section{Discoveries}

Students of Hammersmith Academy discovered four previously unknown variable stars in the Centaurus constellation. 36 observations of 60 seconds each were taken with the T31 telescope centered at RA: 14:51:30, Dec: -35:18:00 on the 30\textsuperscript{th} May 2016. All of the discoveries were W UMa eclipsing variables (contact binaries). Each is composed of a binary system of two stars, each with a higher temperature than the Sun and spectral type A or F. \\

Students of UCL Academy discovered five previously unknown stars in the Libra constellation, with a total of four varieties of variable star from one field of view. Two of the discoveries were W UMa eclipsing variables, with one of them having a shorter than typical period of 0.219 days (5.25 hours), one was a RR Lyrae pulsating variable and another was a flaring red dwarf similar to UV Ceti variables. The fifth was a delta Scuti variable star which has the shortest period, 1.82 hours, of all variable stars discovered in the project and a larger than average amplitude of 1.2 magnitude. \\

A handful of asteroids were found by the pupils within the images but after further investigation and verification they were all found to be already known. Nevertheless, this did not detract from the excitement of finding such objects.\\

The stars which have been accepted into the VSX database at the time of writing are available to view using the links below. Please refer to the web version of the article to access these. 

\begin{flushleft}
Reynolds v1: \url{https://www.aavso.org/vsx/index.php?view=detail.top&oid=473792}

Mousa v1: \url{https://www.aavso.org/vsx/index.php?view=detail.top&oid=474019}

Mousa v2: \url{https://www.aavso.org/vsx/index.php?view=detail.top&oid=473793}

Hamed v1: \url{https://www.aavso.org/vsx/index.php?view=detail.top&oid=474020}

\end{flushleft}

\section{Teachers Feedback}

Feedback forms were filled out by the teachers at each school following every session. The observations made are summarised here.\\

It was noted that IT issues were unavoidable and the need for a back-up session if the software failed became apparent. It was suggested that a member of each school\textsc{\char13}s IT department would be a beneficial future addition to the team.\\

The initial interest from students in the sessions was very high. It was found, however, that attendance dropped off in the middle of the sessions, at one school in particular. Software issues, forgetting the sessions were on and possible frustration were noted as potential reasons for this dip. Attendance increased again after encouragement and weekly reminders about the sessions from the teacher. It was noted that those that did continue to attend throughout were very enthusiastic. Taking this variable attendance into account, it could be beneficial in the future to have a waiting list incase of any drop-outs.\\

It was apparent that the students enjoyed the variety of software offered to them, but that sometimes the difficulty level was too high, due to them being designed for use by professional astronomers. It was agreed that the instructions were far too lengthy and verbose, leading to the need for a lot of one-on-one help from the astronomers and in some cases frustration with the software for not being able to complete the tasks without assistance. Although this was the overall impression, others did seem to enjoy the challenge.\\

Overall the students appeared to enjoy the activities and software related to the discovery of asteroids and variable stars, along with the chance to answer questions and display their prior knowledge of astronomy. The fact that the images given to them were current and used in real research was a positive. There was a slight concern that those who did not make a discovery may feel left out, however this seemed to disappear during the celebration and prize giving closing ceremony, as all students were treated and rewarded equally, regardless of whether they had officially made a new discovery or not.\\

\section{Discussion}

In response to the teacher\textsc{\char13}s feedback regarding the level of the instructions given to the students, the workbooks have now been completely reviewed and revised by UCL\textsc{\char13}s Outreach Officer and the astronomers. It is hoped that the issue regarding the complexity of the software will be mediated by clearer and more straightforward instructions, but it has been noted that in the future it would be beneficial to develop more intuitive versions of the professional astronomy tools that could be used by a wider audience with less technical knowledge. Obviously this would take time and require funding, but it is a potential long term goal with regards to this project. \\

Teaching young students in comparison to interacting with peers who work in this field of research was a challenge for the astronomers but overall it worked well. Although the students needed constant guidance and required instructions written in a clear, step-by-step fashion, they were quick to learn and their familiarity with technology and ability to pick up new software was apparent. Their enthusiasm in having the astronomers available to ask a large variety of questions to was also obvious. \\

The project required a fair amount of preparation time from the astronomers, in the region of a couple of hours a week. This was considered necessary in order for the lessons to run smoothly. Though this time commitment could sometimes be a strain on workload, overall the astronomers felt that the project was "fun, motivating to see the student\textsc{\char13}s enthusiasm and greatly aided with teaching and communication skills". There was no additional time commitment from the teachers and students on top of the hour after school for each weekly session. \\ 

The submission stage of the project was unable to be completed within the time-scale of the clubs for one of the schools, primarily due to the aforementioned software issues. Ultimately, responsibility for submission of these was placed on the astronomers working at each school and flexibility was required in the amount of work that can be covered within the time of the sessions. \\

Feedback from the students involved was taken informally during the closing session, in addition to the aforementioned teachers observations throughout the project. It was found to be positive, with comments such as ``I never knew a complex subject like astronomy would fascinate me greatly, I will now definitely consider it!'' and ``I feel like a real astronomer''. Such feedback implies the club is a promising way to show young students how exciting and rewarding research can be. The students chosen for the project often have little experience of what it might be like to pursue science and a major aim of the project was to show them at an early stage of their education that it is a viable option for their future. There was often a desire to stay longer than the hour allotted time for the club or to continue in their own time. It would be interesting to follow up with the students in a years time to evaluate any potential impact of the project. \\

\section{Conclusions}\label{sec:fin}

Astronomy is a particularly engaging subject for students of this age and one that can incite enthusiasm and active participation in a science subject. It is hoped that this exposure at a young age will encourage those involved to continue to think about science subjects in a positive way and they feel like they can actively be involved in studying them as they continue their education. \\

A total of nine variable star discoveries were made by the participating students during the ten week pilot at Hammersmith Academy and UCL Academy. In order for the project to continue in London, and hopefully branch out to be accessible to a wider range of students and schools, a training scheme is being set up where the astronomers can transfer knowledge of the software and the required skills to interested teachers. Those teachers will then be able to conduct the sessions themselves, in the same format as before. This would make the project more sustainable, as it previously required a large time commitment from local astronomers, including travel time to the schools; something which may be hard to maintain. Some support from astronomers could still be provided, but with the intention that the sessions could be carried out without their presence being strictly necessary. The benefit of the scheme continuing to run at the same school for successive years would be a stronger knowledge of the IT and how to deal with any future issues. It also enables the teacher to have a more active role in the club which they will hopefully find rewarding and a source of enjoyment. \\

\section*{Acknowledgements}

We would like to express our gratitude to teachers Harriet Silcocks at Hammersmith Academy and Colin Ankerson at UCL Academy who readily gave their time to stay after school to supervise and assist in the running of these sessions. To our collaborators in Russia, Denis Denisenko and Stanislav
Korotkiy at Sternberg Astronomical Institute, Lomonosov Moscow State University for training and supporting us throughout the project. This project was made possible by the Noosfera Education Support Foundation and Mayor\textsc{\char13}s Fund for London. The initiative is also supported by the BE OPEN foundation. Tess Sullivan and Carole Coulon from the Mayor\textsc{\char13}s fund and Maria Budilina and Sofia Arzumanyan from Noosfera were involved throughout. \\

\bibliographystyle{unsrt}
\bibliography{ed_ast} 

\end{document}